\documentclass[twoside]{article}
\input oejv.sty

\usepackage[version=4]{mhchem}
\usepackage{amsmath,amssymb}

\begin{document}

\OEJVhead{November 2022}
\OEJVtitle{UBVRI photometry of Betelgeuse over 23 years since 1999}
\OEJVauth{Yojiro Ogane$^{1}$, Osamu Ohshima$^{2,3}$, Daisuke Taniguchi$^{4}$, and Naohiro Takanashi$^{5}$}
\OEJVinst{Ogane Hikari Observatory, 3-28-29 Hikari, Kokubunji, Tokyo 185-0034, Japan, {\tt \href{mailto:email}{hqk00742@nifty.com}}}
\OEJVinst{Okayama University of Science, 1-1 Ridaicho, Kita-ku, Okayama, Okayama 700-0005, Japan}
\OEJVinst{Ohshima Tamashima Observatory, 3-10-15 Tamashima, Kurashiki, Okayama 713-8102, Japan, {\tt \href{mailto:email}{o2@otobs.org}}}
\OEJVinst{Department of Astronomy, The University of Tokyo, 7-3-1 Hongo, Bunkyo-ku, Tokyo 113-0033, Japan, {\tt \href{mailto:email}{d.taniguchi.astro@gmail.com}}}
\OEJVinst{Executive Management Program, The University of Tokyo, 7-3-1 Hongo, Bunkyo-ku, Tokyo 113-8654, Japan, {\tt \href{mailto:email}{naohiro.takanashi@emp.u-tokyo.ac.jp}}}

\OEJVabstract{
We report the results of our continuous \textit{\fontsize{10pt}{0pt}\selectfont UBVRI}-band photometry of Betelgeuse from 1999 to 2022 using the same photometric system. 
There are two advantages in our observation: (1)~we used a photodiode as a detector to avoid saturation, and (2)~our data set includes \textit{\fontsize{10pt}{0pt}\selectfont U}-band light curve, which is not widely observed in recent CCD photometries. 
Using our light curves, we conducted the periodicity analysis, and found ${\sim }$405- and ${\sim }$2160-day periods. 
We also discuss the tentative detection of a long-period variation over 20 years or longer. 
Finally, we discuss the peculiar variation of the $U-B$ color index during the ``Great Dimming'' event between late 2019 and early 2020.
}

\begintext

\section{Introduction}

Located at the Orion's shoulder, Betelgeuse is one of the nearest red supergiant stars, and is also known as a semi-regular variable~\citep{lloyd2020}. 
In 1980s, when this project was initiated, there was an active discussion on the variable period(s) of Betelgeuse~\citep[see][and references therein]{dupree1987}, and an emergence of a long-term, homogeneous photometric dataset was anticipated. 
Under such circumstance, we started to investigate the photometric periodicity of Betelgeuse by performing accurate, multi-color photometry with the same photometric system over tens of years. 

Nowadays, thanks to many variable-star observers worldwide contributing to the American Association of Variable Star Observers (AAVSO) database, the variable periods of Betelgeuse in the visible band has been well measured: e.g., $185{\pm }13.5$, $416{\pm }24$, and $2365{\pm }10\,\mathrm{days}$~\citep{joyce2020}. 
Even in such an era, a long-term observing campaign by a specific observer would be valuable, considering that such data set like the AAVSO database cannot, by definition, circumvent systematic biases among the photometric data provided by multiple observers. 

In addition, starting from the latter half of 2019, so-called ``Great Dimming'' event occurred~\citep{guinan2020,dupree2022}, and many well-established telescopes pointed towards Betelgeuse. 
However, the cause of the Dimming is still under debate~\citep{montarges2021}. 
Proposed explanations include, for example, a decrease in the effective temperature~\citep[e.g.][]{harper2020}, an increase in the (circumstellar) dust extinction~\citep[e.g.][]{dupree2020}, and both of them~\citep[e.g.][]{levesque2020}. 
One can also interpret the observational evidences as consequences of a surface mass ejection~\citep{dupree2022}. 

In this paper, we present our \textit{UBVRI} multi-band photometry of Betelgeuse over $23$ years, and determine the variable periods for the \textit{V} band. 
Moreover, with the aid of the long-term, multi-band nature of our dataset, especially the inclusion of the \textit{U} band, we discuss a possible cause of the Great Dimming.

\section{Observation}

\subsection{Observation and basic reduction}

We have been observing Betelgeuse in the \textit{UBVRI} bands since 1999 at the private Ogane Hikari Observatory located in Hikari-cho, Kokubunji City, Tokyo, Japan ($139.4^{\circ }$E, $35.7^{\circ }$N). 
In total, we here present data for $572$ nights, including $17$ nights in which we observed Betelgeuse twice per night. 
We used a $25\,\mathrm{cm}$-aperture, f/8 Newtonian telescope until February 2006, and used a $30\,\mathrm{cm}$-aperture, f/16 Cassegrain telescope after that. 
In order to avoid saturation for very bright stars like Betelgeuse, we used a photodiode (Hamamatsu S1226-5BQ), instead of photomultiplier, as a detector to conduct a single-channel, photoelectric observation. 

\begin{figure}[t]
	\centering
	\includegraphics[width=12cm]{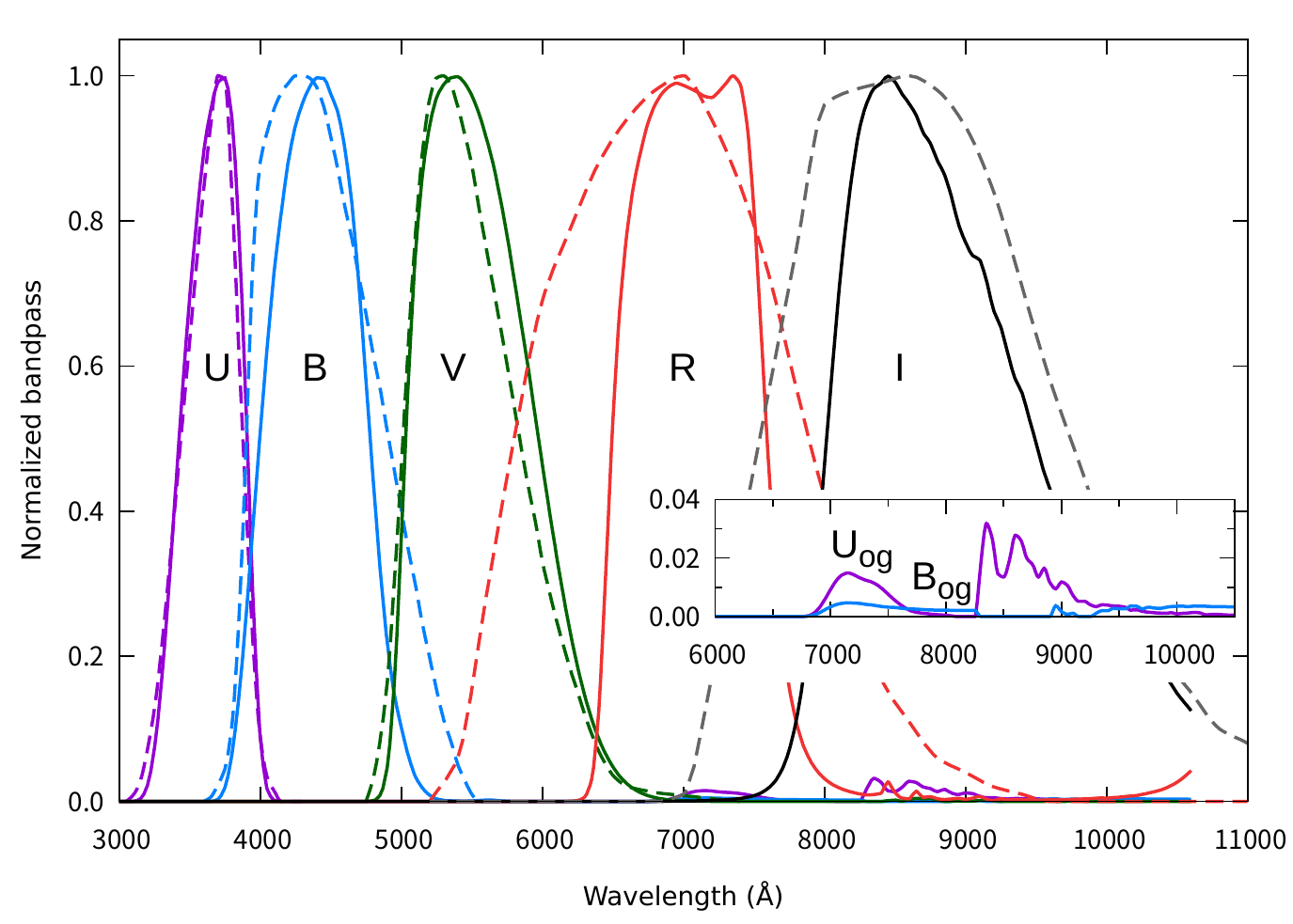}
	\caption{Normalized bandpass of \textit{UBVRI} systems. Solid and dashed lines represent Ogane's and standard \citet{johnson1965} systems, respectively. In the long wavelength side ($\lambda >7000\,\text{\AA }$), there are small red leakage to the \textit{U} and \textit{B} bands, which has a non-negligible effect to the photometry of Betelgeuse (see the main text). }
	\label{fig1}
\end{figure}

To build a photometric system imitating the standard Johnson's \textit{UBVRI} system~\citep{johnson1965}, we stacked a few commercially-available colored-glass filters for each band. 
The resultant bandpass for each band, i.e. the combination of the photodiode's sensitivity and filters' throughputs, is showed in \autoref{fig1} and \autoref{tab1}. 
There are some discrepancies between the Ogane's and Johnson's system. 
For example, the effective wavelength of the Ogane's \textit{B} band is slightly longer than the Johnson's one, and the band width is narrower. 
Because of these discrepancies in the photometric bands, we distinguish the Ogane's system from the standard Johnson's system by adding ``og'' subscripts, if necessary.

\begin{table}[t]
	\caption{\textit{UBVRI} bandpass of the Ogane's photometric system. Full table is available on Appendix (\autoref{tabA1-01}), the OEJVS's web site (\url{https://oejv.physics.muni.cz/issues/table1_0233.txt}), and also on the authors' web site (\url{http://hikariao.la.coocan.jp/figure.html}). }\vspace{3mm}
	\centering
	\begin{tabular}{lccccc}
		\hline
		Wavelength [\AA ] & \textit{U} & \textit{B} & \textit{V} & \textit{R} & \textit{I}  \\ \hline \hline
		3050&0.0000 &0.0000 &0.0000 &0.0000 &0.0000\\
		3100&0.0001 &0.0000 &0.0000 &0.0000 &0.0000\\
		3150&0.0036 &0.0000 &0.0000 &0.0000 &0.0000\\
		3200&0.0218 &0.0000 &0.0000 &0.0000 &0.0000\\
		...&&&&&\\  
		\hline
	\end{tabular}\label{tab1}
\end{table}

We used the all-sky photometry method~\citep{henden1982} for the photometry. 
The differential photometry method cannot be used for our case because no star around Betelgeuse is suitable as a comparison star in terms of its brightness and color. 
As the candidate standard stars for the all-sky photometry, we selected $83$ stars  from the Arizona-Tonantzintla catalog~\citep{iriarte1965}, with the main criteria being (1) non-variable stars and (2) the \textit{V} magnitude of $4.5$ or brighter. 
In each night, $10\text{--}12$ stars among them with the altitudes higher than $30^{\circ }$ were observed. 

Each object among photometric standard stars and Betelgeuse was observed typically once per night for each band. 
The typical integration time for the standards was $80\text{--}100\,\mathrm{s}$ for each of the \textit{BVRI} bands and $200\text{--}400\,\mathrm{s}$ for the less sensitive \textit{U} band. 
For Betelgeuse, which is brighter than most of the standards, we used $80\text{--}90\,\mathrm{s}$ integration for all the bands. 
In total, one set of the observations of standard stars and Betelgeuse took about $5\,\mathrm{hours}$. 
Due to this long time of integrations required for each night, photometric errors because larger when the weather was unstable. 

For each night's data set, we performed a multivariate analysis to determine the extinction coefficients of the telluric atmosphere and transformation coefficients from the natural- to the standard-system \textit{V} magnitude and colors. 
Since our list of standard stars does not include stars as red as Betelgeuse, we extrapolated the relation between observed raw counts and the magnitudes calibrated with the standard system before applying to Betelgeuse. 
This extrapolation could introduce systematic bias/errors to the measured photometric magnitudes, but we found no detectable systematic errors that depend on the color or airmass of an object. 
Note that we removed seven observations with poor weather conditions from our published photometric catalog.

\subsection{Correction for red leakage}

A small portion of red and near-infrared photons passes through the blue (\textit{U}$_{\mathrm{og}}$ and \textit{B}$_{\mathrm{og}}$) filters, as shown in \autoref{fig1}. 
This red leakage has a non-negligible effect to our photometry, especially for M-type stars; for example, the uncorrected \textit{U}$_{\mathrm{og}}$-band flux of Betelgeuse is contributed almost the same amounts by \textit{U}-band and \textit{RI}-band photons. 

\begin{figure}[t]
	\centering
	\includegraphics[width=14cm]{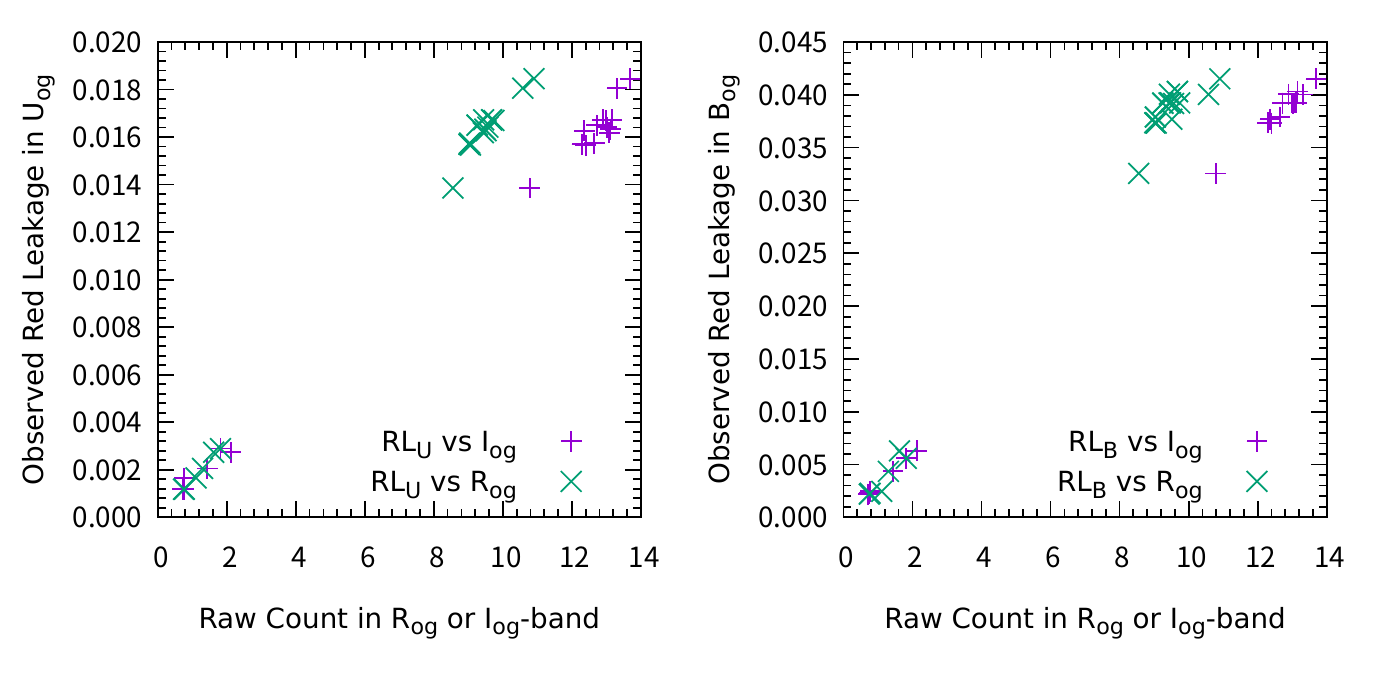}
	\caption{Red leakage in \textit{U}$_{\mathrm{og}}$ (left panel) and \textit{B}$_{\mathrm{og}}$ (right panel) bands vs \textit{R}$_{\mathrm{og}}$ (green cross) and \textit{I}$_{\mathrm{og}}$ (purple plus) raw counts. Data points at the top right corners, i.e. those with the $x$-axis values larger than $8$, show the results for Betelgeuse observed on different nights, and are used for the fitting. }
	\label{fig2}
\end{figure}

In order to investigate the effect of the red leakage to our photometry, we measured, in summer 2021, the amounts of raw \textit{U}$_{\mathrm{og}}$\textit{B}$_{\mathrm{og}}$ counts contributed by red photons. 
For this purpose, we inserted a red filter transparent at wavelengths longer than $6500\,\text{\AA }$ into each of the \textit{U}$_{\mathrm{og}}$- and \textit{B}$_{\mathrm{og}}$-band filter systems. 
Then, we observed Betelgeuse for multiple times and also some other red stars with these stacked filters. 
\autoref{fig2} shows the correlations between the amount of the red leakage and the amount of the raw count in the \textit{R}$_{\mathrm{og}}$ and \textit{I}$_{\mathrm{og}}$ bands. 
We fitted these data, only of Betelgeuse, using linear regression, and obtained the following equations to best correct for the red leakage: 

\begin{align}
   B_{\mathrm{count}} = B_{\mathrm{raw}} - ((3.27\times 10^{-3})I_{\mathrm{raw}} - (2.38\times 10^{-4})R_{\mathrm{raw}} - 7.44\times 10^{-4})/0.895, \label{eq1} \\
   U_{\mathrm{count}} = U_{\mathrm{raw}} - ((7.63\times 10^{-4})I_{\mathrm{raw}} + (9.97\times 10^{-4})R_{\mathrm{raw}} - 2.77\times 10^{-4})/0.895. \label{eq2}
\end{align}    

Here, the subscript ``raw'' indicates the raw count for each band before the atmospheric correction, and ``count'' indicates the counts after correcting the red leakage. 
The constant $0.895$ is the transmittance at red wavelengths of the inserted red filter. 
These equations are related to the ratios of the throughputs between the \textit{RI} and \textit{UB} filters in red wavelength ranges, and thus expected to be independent of the used telescopes. 
In other words, although these equations are determined using the new $30\,\mathrm{cm}$ telescope, they are expected to be applicable to the photometric data obtained with the old $25\,\mathrm{cm}$ telescope.

We corrected \textit{U}$_{\mathrm{og}}$ and \textit{B}$_{\mathrm{og}}$ data observed before summer 2021 using the \textit{R}$_{\mathrm{og}}$\textit{I}$_{\mathrm{og}}$-band counts using these equations. 
For the observations after autumn 2021, we measured and subtracted the red leakage in \textit{U}$_{\mathrm{og}}$ and \textit{B}$_{\mathrm{og}}$ using the red filter. 
Note that even after the correction for the red leakage by this procedure and the transformation to the standard system, we cannot eliminate a possible systematic bias of our \textit{U}$_{\mathrm{og}}$\textit{B}$_{\mathrm{og}}$ photometry compared to the Johnson's system due to, e.g., the different filter transmissions and an imperfect correction of the red leakage.

\section{Results and discussion}

\subsection{Time variation of the magnitudes and color indices}

\begin{figure}[t]
	\centering
	\includegraphics[width=10cm]{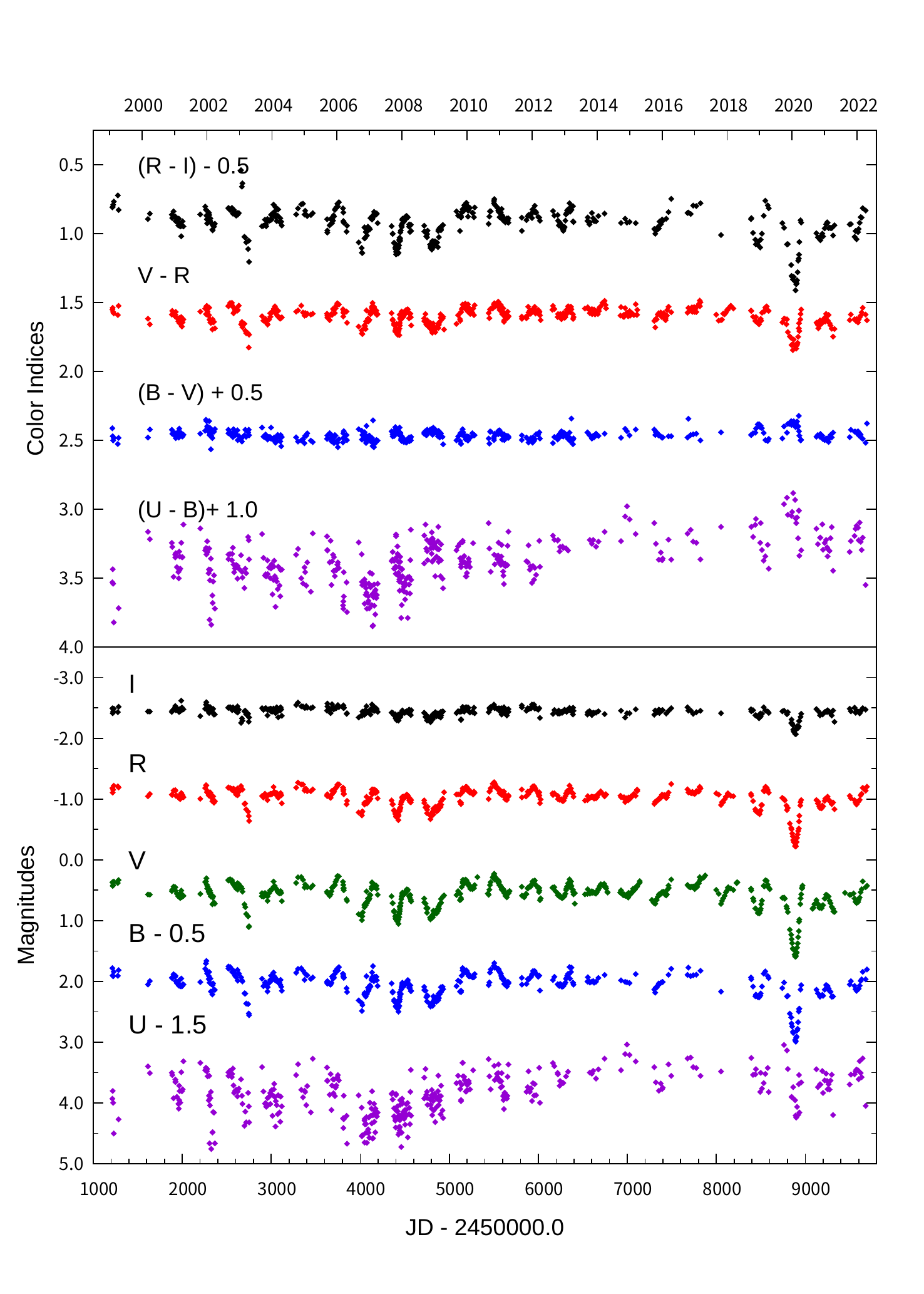}
	\caption{Magnitude and color variations of Betelgeuse from 1999 to 2022. Upper panel: color indices of $U - B$, $B - V$ ,$V - R$, and $R - I$. Lower panel: \textit{UBVRI} magnitudes. }
	\label{fig3}
\end{figure}

The resultant magnitudes and color indices of Betelgeuse for each night are shown in \autoref{fig3} and  \autoref{tab2}. 
The ``err'' columns in \autoref{tab2} show the standard deviation of the residuals around the multivariate regressions for correcting the atmospheric extinction and photometric-system differences using the standard stars' data. 

\begin{figure}[t]
	\centering
	\includegraphics[width=10cm]{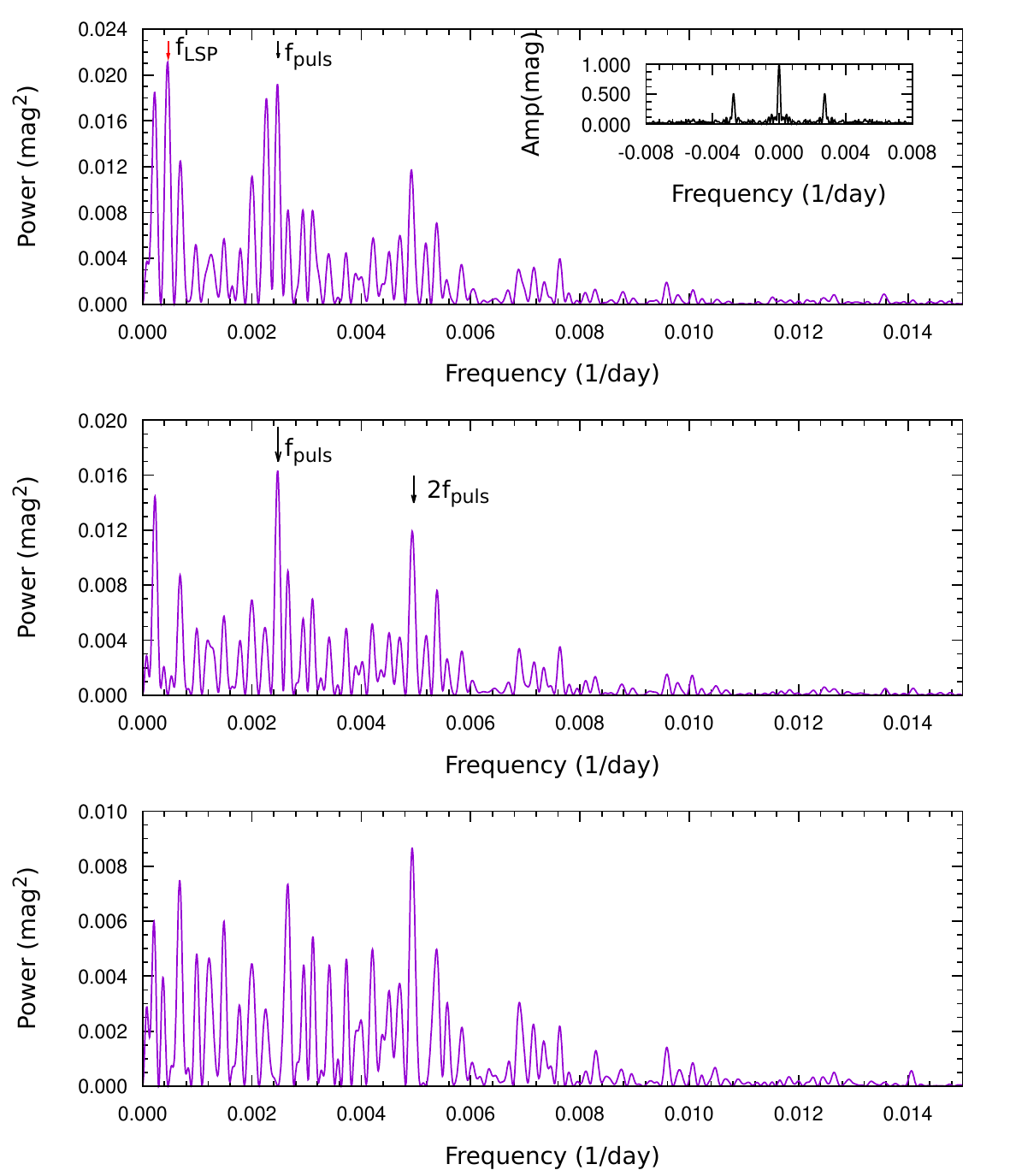}
	\caption{The top panel shows the power spectrum of our \textit{V}-band light curve. The strongest frequency, i.e., the long-secondary period $f_{\mathrm{LSP}}$ (LSP), is indicated by the red arrow. The inset shows the spectral window function of the data. The middle panel shows the power spectrum after removing the LSP signals. The peaks corresponding to the fundamental model pulsation ($f_{\mathrm{puls}}$) and its first harmonic (2$f_{\mathrm{puls}}$) are indicated by the black arrows. The bottom panel shows the residual spectrum after removing the aforementioned frequencies. }
	\label{fig4}
\end{figure}

The periodicity in the obtained \textit{V}-band light curve was analyzed using \textsc{Period04} code\footnote{\url{https://univie.ac.at/tops/Period04/}}~\citep{lenz2005}. 
We detected mainly three period components: ${\sim }2160$, ${\sim }405$, and ${\sim }202\,\mathrm{days}$, probably corresponding to the long secondary period (LSP), fundamental mode pulsation, and first harmonics of the fundamental mode, respectively (\autoref{fig4}). 
The errors in measuring peak frequencies in the power are several days for the longest period and smaller than one day for the others. 
Although these measurement errors would be smaller than the uncertainties of ``physical'' periods~\citep{kiss2006}, our determined periods are within, or at least near to, the ranges of previous measurements by \citet{kiss2006,joyce2020}, which supports the reliability of our period determination.

\begin{table}[t]
	\caption{\textit{UBVRI} photometry of Betelgeuse by Ogane. Full table is available on Appendix (\autoref{tabA2-01}), the OEJVS's web site (\url{https://oejv.physics.muni.cz/issues/table2_0233.txt}), and also on the authors' web site (\url{http://hikariao.la.coocan.jp/data.html}). }\vspace{3mm}
	\centering
	\begin{tabular}{lcccccccccc}
		\hline
		Hel.J.D. &\textit{V} & err&$U - B$&err&$B -V$&err&$V - R$&err&$R - I$&err \\ \hline \hline
		2451215.04 &0.369 &0.043 &2.531 &0.072 &1.912 &0.037 &1.551 &0.022 &1.309 &0.014 \\
		2451217.98 &0.431 &0.041 &2.436 &0.042 &1.970 &0.017 &1.538 &0.013 &1.306 &0.016 \\
		2451224.05 &0.416 &0.013 &2.541 &0.025 &2.001 &0.007 &1.567 &0.010 &1.291 &0.018 \\
		2451231.05 &0.359 &0.011 &2.822 &0.120 &1.984 &0.014 &1.578 &0.013 &1.267 &0.013 \\
		2451275.98 &0.382 &0.010 &      &      &2.026 &0.010 &1.590 &0.008 &1.222 &0.021 \\
		2451284.96 &0.334 &0.015 &2.718 &0.111 &1.982 &0.021 &1.524 &0.020 &1.328 &0.036 \\
		2451613.96 &0.571 &0.020 &2.164 &0.068 &1.980 &0.027 &1.617 &0.025 &1.394 &0.017 \\
		2451635.97 &0.574 &0.015 &2.218 &0.099 &1.921 &0.033 &1.658 &0.016 &1.355 &0.028 \\
		2451883.15 &0.515 &0.013 &2.244 &0.026 &1.924 &0.011 &1.587 &0.017 &1.362 &0.017 \\
		2451888.15 &0.494 &0.019 &2.274 &0.035 &1.953 &0.007 &1.560 &0.018 &1.386 &0.015 \\
		...&&&&&&&&&\\  
		\hline
	\end{tabular}\label{tab2}
\end{table}

\begin{figure}[t]
	\centering
	\includegraphics[width=14cm]{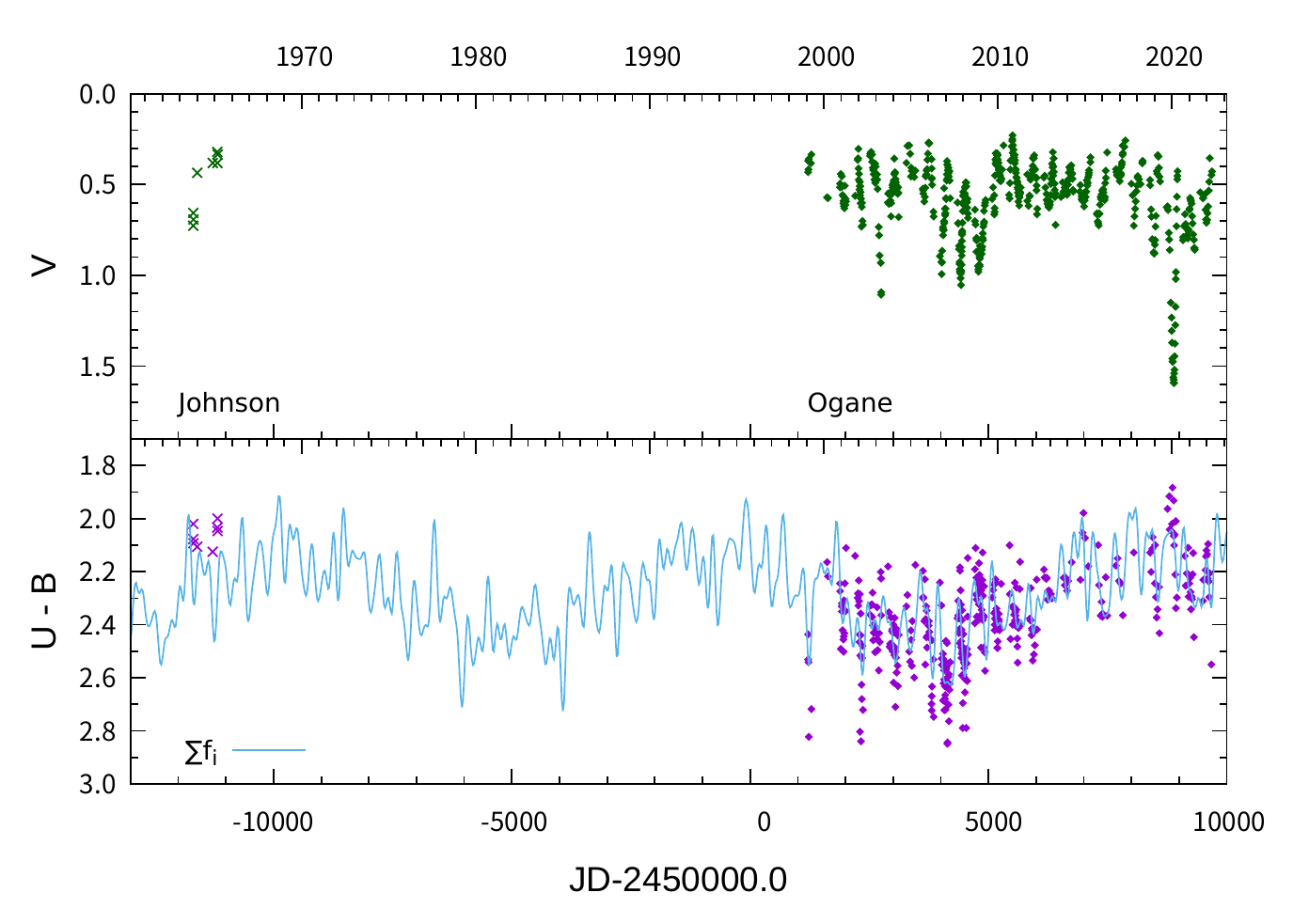}
	\caption{Long-term \textit{V} and $U - B$ variation of Betelgeuse. Dots and crosses represent the data measured by this work and by \citet{johnson1966}, respectively. There are two type of variations: the pulsation with the ${\sim }400$-day period and long-term variation with the time scale of a few tens of years. Blue thin line in the lower panel shows the synthesized light curve using top ten Fourier components of the Ogane's data points; we do not include \citet{johnson1966} data points in this Fourier transformation. }
	\label{fig5}
\end{figure}

We have also examined the long-term variation of Betelgeuse, especially in the $U - B$ color index. 
\citet{johnson1966} reported multi-band photometry of Betelgeuse during the 2-year duration starting from $\mathrm{JD}=2438315$ (i.e. October 1963). 
For the \textit{V} magnitude and $U - B$ color, they observed Betelgeuse for eight epochs. 
There is a gap of about $35$ years between the Johnson's observations and ours, and we could not find any published $U - B$ measurements during this gap.
Plotting Johnson's and our $U - B$ data (lower panel of \autoref{fig5}), we found a long-term variation in the $U-B$ color, possibly with a period of ${\sim }27\,\mathrm{years}$. 
Such long-timescale variation is also seen in the AAVSO's \textit{V}-band light curve~\citep[e.g.][]{montarges2021}, and it is possibly due to the rotational modulation~\citep{joyce2020} considering the rotational period of ${\sim }31\,\mathrm{years}$~\citep{kervella2018}. 
Thus, the long-term $U-B$ variation that we found might be attributed to an actual variation of Betelgeuse over the $35$-year gap, at least in part, rather than the systematic bias and/or errors. 
This is also supported by the fact that the synthesized and extrapolated $U-B$ light curve using only Ogane's data (blue thin line in the lower panel of \autoref{fig5}) well predicts the \citet{johnson1966} data points. 
In addition, we found that the \textit{V}-band variation (upper panel of \autoref{fig5}) is weaker than the $U - B$ color one. 
Nevertheless, possible systematic bias between the two observers, i.e., we and \citet{johnson1966}, should be kept in mind. 
Future longer-term, continuous observations will be required to further examine these longer period phenomena.

\subsection{Color-magnitude diagram: peculiar behavior of the $U-B$ color during the Great Dimming}

\begin{figure}[t]
	\centering
	\includegraphics[width=10cm]{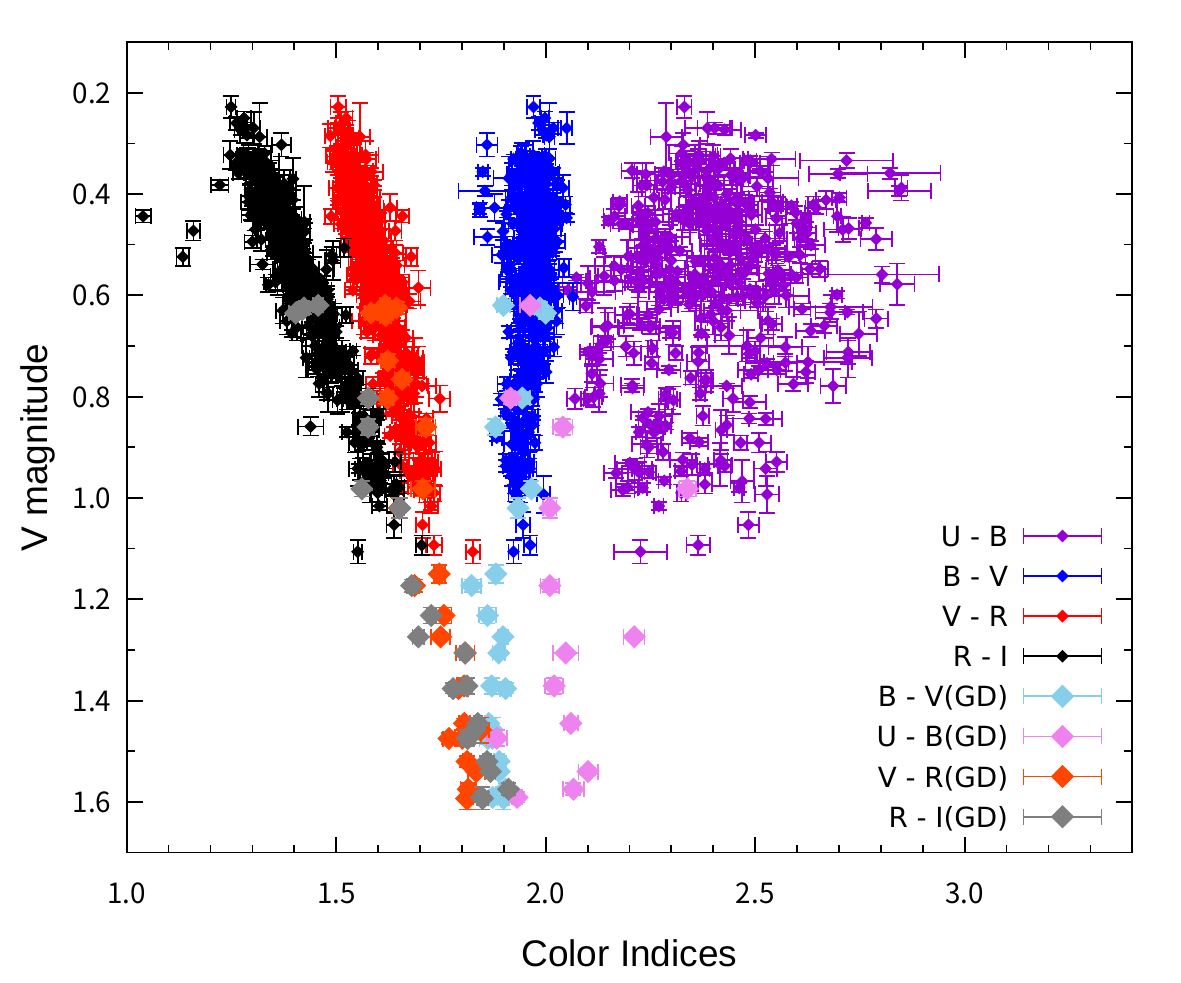}
	\caption{Color-magnitude diagrams of our photometric data set. Small and large symbols represent the data during and outside the Great Dimming, respectively. }
	\label{fig6}
\end{figure}

\begin{figure}[t]
	\centering
	\includegraphics[width=16cm]{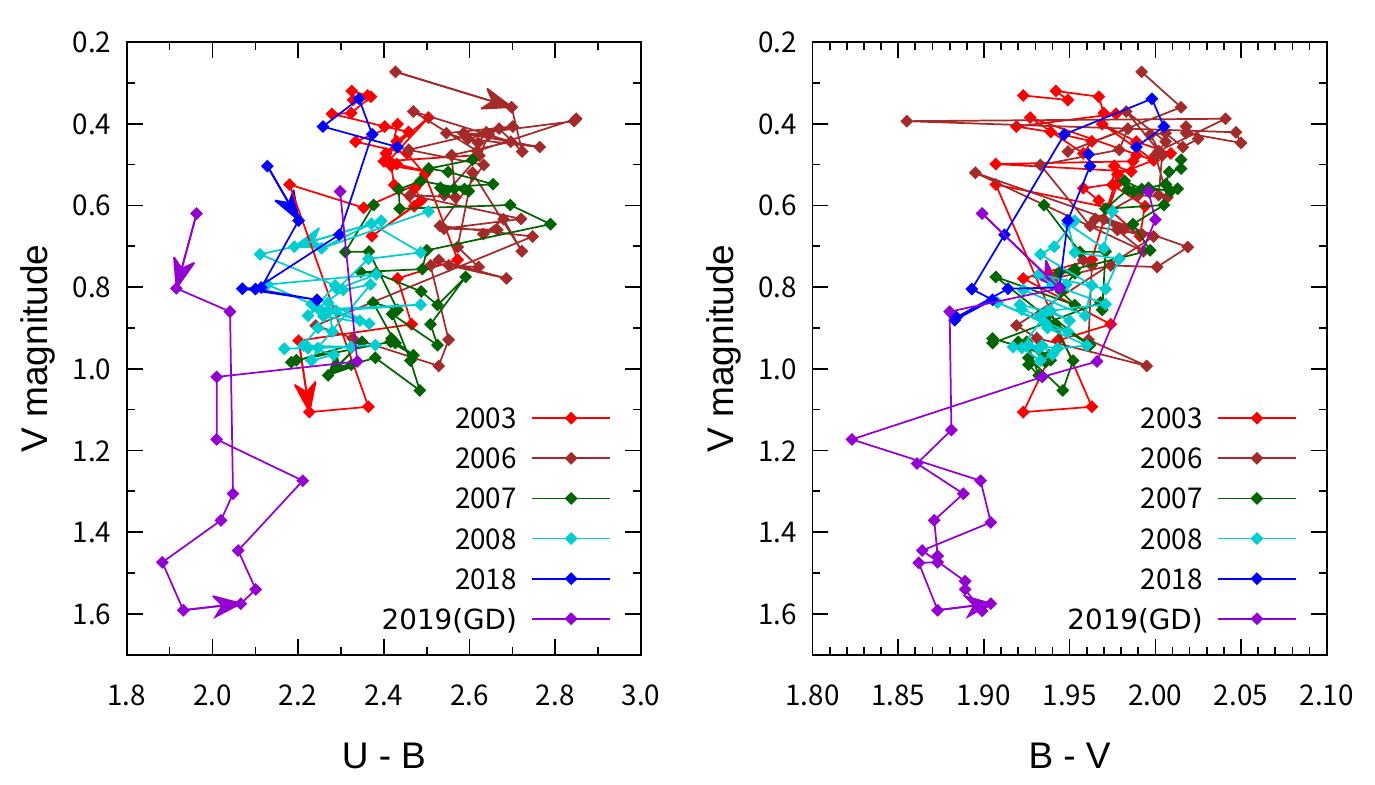}
	\caption{The color-magnitude diagram  between the \textit{V} magnitude and $U-B$ color index for some variability cycles, including the Great Dimming (GD) phase in 2019. }
	\label{fig7}
\end{figure}

During the usual phases of Betelgeuse, i.e. excluding the Great Dimming phase from late 2019 to early 2020, the color indices and the \textit{V} magnitude that we measured exhibit good correlations, especially for the $B-V$, $V-R$, and $R-I$; when the \textit{V} magnitude is lower than the average, $V-R$ and $R-I$ colors are redder and $U-B$ and $B-V$ colors are bluer (\autoref{fig6}). 
These correlations could be mainly attributed to the variation in the effective temperature, considering the dependence on the temperature of the black body shape and of the strengths of the \ce{TiO} molecular bands in the \textit{BVR} bands~\citep{mckemmish2019}, latter of which are very sensitive to the effective temperature~\citep{levesque2005}. 

The scatter around the relation between \textit{V} and $U-B$ is larger than the observational error for $U-B$ quoted in \autoref{tab2}. 
To reveal the origin of this scatter, we plotted the long-term change in the trajectory between $U - B$ and \textit{V} of the individual $400$-day cycle (\autoref{fig7}). 
As a result, we found differences in the trajectories between different cycles. 
For example, in the 2003 cycle (i.e. late 2003--early 2004), which is the second largest dimming event during our observation after the Great Dimming, the average color index is $U-B\sim 2.5$. 
In contrast, during the Great Dimming period, the average color index is bluer, $U-B\sim 2.0$. 
These differences suggest that not only the effective temperature and dust extinction but also the chromospheric activity~\citep{davies2021} may affect the $U-B$ color. 
Therefore, the large scatter might not be observational systematic errors, but rather than the actual variation of the spectral-energy distribution of Betelgeuse. 

Focusing onto the Great Dimming, the trajectory on the color-magnitude diagram between $U - B$ and \textit{V} exhibited unusual behavior: the color remained mostly unchanged, or even became redder when the \textit{V}-magnitude is fainter (right panel of \autoref{fig7}). 
Considering that the temperature was decreased by ${\sim }100\,\mathrm{K}$ during the Dimming~\citep[e.g.][]{levesque2020,taniguchi2022}, which makes the $U-B$ color ${\sim }0.2\,\mathrm{mag}$ bluer, effect(s) that makes the color redder should have happened. 
Plausible scenarios include the increased circumstellar extinction~\citep{montarges2021,taniguchi2022}. 

Moreover, when the \textit{V} magnitude was staying at the minimum of ${\sim }1.6\,\mathrm{mag}$ on February 2020, $U-B$ became redder and $B-V$ did slightly bluer by ${\sim }0.2$ and ${\lesssim }0.1\,\mathrm{mag}$, respectively. 
These variations in the colors, without variation in the \textit{V} magnitude, could be explained, for example, assuming that the following three phenomena occurred in succession. 
First, the chromosphere experienced an increased activity during the second half of 2019, i.e., only before the Great Dimming~\citep{dupree2020}. 
Second, the dust extinction was enhanced for about half a year, and it reached its maximum before the Dimming on February 2020~\citep{taniguchi2022}. 
Finally, the surface temperature was cooler than the usual for about half a year, and it reached its minimum after the Dimming~\citep{kravchenko2021}.

\setcounter{secnumdepth}{0}
\OEJVacknowledgements{
We acknowledge useful comments from the referee. 
This work was supported by MEXT KAKENHI Grant Numbers 06916011 and 11916014. 
This work was also supported by the Support Fund for Domestic Training Programs in Astronomy (the Naichi-Ryugaku Scholarship) from the Astronomical Society of Japan. 
DT acknowledges financial support from Masason Foundation. 
We are grateful Prof. Wataru Tanaka at the National Astronomical Observatory of Japan for supervising this project since 1993. 
We thank Mr. Kikuichi Arai at North Riverside Observatory for his supports in writing this manuscript. 
We also thank Dr. Noriyuki Matsunaga and Dr. Miguel Montarg\`es for their helpful discussion regarding the Great Dimming. } \\

\renewcommand{\thetable}{A1}

\begin{table}[t]
	\caption{Full table of \autoref{tab1}: \textit{UBVRI} bandpass of the Ogane's photometric system. }\vspace{3mm}
	\centering
\begin{tabular}{lccccc}
		\hline
		Wavelength [\AA ] & \textit{U} & \textit{B} & \textit{V} & \textit{R} & \textit{I}  \\ \hline \hline
		3000&0.0000 &0.0000 &0.0000 &0.0000 &0.0000\\
		3050&0.0000 &0.0000 &0.0000 &0.0000 &0.0000\\
		3100&0.0001 &0.0000 &0.0000 &0.0000 &0.0000\\
		3150&0.0036 &0.0000 &0.0000 &0.0000 &0.0000\\
		3200&0.0218 &0.0000 &0.0000 &0.0000 &0.0000\\
		3250&0.0732 &0.0000 &0.0000 &0.0000 &0.0000\\
		3300&0.1641 &0.0000 &0.0000 &0.0000 &0.0000\\
\hline
	\end{tabular}\label{tabA1-01}
\end{table}

\begin{table}[t]
	\caption{Continued. }\vspace{3mm}
	\centering
	\begin{tabular}{lccccc}
		\hline
		Wavelength [\AA ] & \textit{U} & \textit{B} & \textit{V} & \textit{R} & \textit{I}  \\ \hline \hline
		5000&0.0000 &0.1024 &0.3623 &0.0000 &0.0000\\
		5050&0.0000 &0.0631 &0.5788 &0.0000 &0.0000\\
		5100&0.0000 &0.0344 &0.7516 &0.0000 &0.0000\\
		5150&0.0000 &0.0184 &0.8640 &0.0000 &0.0000\\
		5200&0.0000 &0.0087 &0.9302 &0.0000 &0.0000\\
		5250&0.0000 &0.0036 &0.9697 &0.0000 &0.0000\\
		5300&0.0000 &0.0016 &0.9877 &0.0000 &0.0000\\
\hline
	\end{tabular}\label{tabA1-02}
\end{table}

\begin{table}[t]
	\caption{Continued. }\vspace{3mm}
	\centering
	\begin{tabular}{lccccc}
		\hline
		Wavelength [\AA ] & \textit{U} & \textit{B} & \textit{V} & \textit{R} & \textit{I}  \\ \hline \hline
		7000&0.0092 &0.0033 &0.0009 &0.9867 &0.0005\\
		7050&0.0123 &0.0041 &0.0005 &0.9824 &0.0006\\
		7100&0.0142 &0.0045 &0.0003 &0.9774 &0.0008\\
		7150&0.0149 &0.0047 &0.0002 &0.9718 &0.0010\\
		7200&0.0143 &0.0046 &0.0001 &0.9699 &0.0014\\
		7250&0.0133 &0.0045 &0.0000 &0.9780 &0.0017\\
		7300&0.0125 &0.0042 &0.0000 &0.9891 &0.0022\\
\hline
	\end{tabular}\label{tabA1-03}
\end{table}

\begin{table}[t]
	\caption{Continued. }\vspace{3mm}
	\centering
	\begin{tabular}{lccccc}
		\hline
		Wavelength [\AA ] & \textit{U} & \textit{B} & \textit{V} & \textit{R} & \textit{I}  \\ \hline \hline
		9000&0.0118 &0.0017 &0.0002 &0.0030 &0.7707\\
		9050&0.0107 &0.0008 &0.0004 &0.0045 &0.7525\\
		9100&0.0072 &0.0011 &0.0013 &0.0069 &0.7463\\
		9150&0.0052 &0.0000 &0.0016 &0.0029 &0.7163\\
		9200&0.0052 &0.0000 &0.0010 &0.0011 &0.6765\\
		9250&0.0040 &0.0000 &0.0000 &0.0037 &0.6539\\
		9300&0.0034 &0.0011 &0.0001 &0.0020 &0.6175\\
		9350&0.0040 &0.0020 &0.0001 &0.0028 &0.5836\\
		9400&0.0039 &0.0020 &0.0000 &0.0026 &0.5631\\
		9450&0.0036 &0.0028 &0.0001 &0.0020 &0.5289\\
		9500&0.0035 &0.0026 &0.0000 &0.0020 &0.4941\\
		9550&0.0032 &0.0026 &0.0000 &0.0019 &0.4582\\
		9600&0.0022 &0.0036 &0.0004 &0.0024 &0.4299\\
		9650&0.0023 &0.0035 &0.0001 &0.0026 &0.4049\\
		9700&0.0018 &0.0026 &0.0001 &0.0021 &0.3830\\
		9750&0.0015 &0.0032 &0.0003 &0.0028 &0.3633\\
		9800&0.0013 &0.0029 &0.0003 &0.0033 &0.3391\\
		9850&0.0012 &0.0028 &0.0000 &0.0035 &0.3207\\
		9900&0.0011 &0.0031 &0.0000 &0.0041 &0.3058\\
		9950&0.0013 &0.0034 &0.0001 &0.0047 &0.2981\\
		10000&0.0009 &0.0035 &0.0000 &0.0046 &0.2913\\
		10050&0.0011 &0.0035 &0.0000 &0.0053 &0.2789\\
		10100&0.0012 &0.0034 &0.0000 &0.0060 &0.2612\\
		10150&0.0014 &0.0036 &0.0001 &0.0071 &0.2428\\
		10200&0.0012 &0.0035 &0.0001 &0.0078 &0.2307\\
		10250&0.0006 &0.0035 &0.0001 &0.0096 &0.2220\\
		10300&0.0007 &0.0034 &0.0001 &0.0117 &0.2067\\
		10350&0.0007 &0.0034 &0.0002 &0.0139 &0.1913\\
		10400&0.0006 &0.0034 &0.0002 &0.0169 &0.1764\\
		10450&0.0005 &0.0033 &0.0001 &0.0212 &0.1640\\
		10500&0.0005 &0.0033 &0.0001 &0.0265 &0.1489\\
		10550&0.0008 &0.0031 &0.0001 &0.0337 &0.1368\\
		10600&0.0007 &0.0030 &0.0001 &0.0425 &0.1249\\
		\hline
	\end{tabular}\label{tabA1-04}
\end{table}

\renewcommand{\thetable}{A2}

\begin{table}[t]
	\caption{Full table of \autoref{tab2}: \textit{UBVRI} photometry of Betelgeuse by Ogane. }\vspace{3mm}
	\centering
	\begin{tabular}{lcccccccccc}
		\hline
		Hel.J.D. &\textit{V} & err&$U - B$&err&$B -V$&err&$V - R$&err&$R - I$&err \\ \hline \hline
		2451215.04 &0.369 &0.043 &2.531 &0.072 &1.912 &0.037 &1.551 &0.022 &1.309 &0.014 \\
		2451217.98 &0.431 &0.041 &2.436 &0.042 &1.970 &0.017 &1.538 &0.013 &1.306 &0.016 \\
		2451224.05 &0.416 &0.013 &2.541 &0.025 &2.001 &0.007 &1.567 &0.010 &1.291 &0.018 \\
		2451231.05 &0.359 &0.011 &2.822 &0.120 &1.984 &0.014 &1.578 &0.013 &1.267 &0.013 \\
		2451275.98 &0.382 &0.010 &      &      &2.026 &0.010 &1.590 &0.008 &1.222 &0.021 \\
		2451284.96 &0.334 &0.015 &2.718 &0.111 &1.982 &0.021 &1.524 &0.020 &1.328 &0.036 \\
		2451613.96 &0.571 &0.020 &2.164 &0.068 &1.980 &0.027 &1.617 &0.025 &1.394 &0.017 \\
\hline
	\end{tabular}\label{tabA2-01}
\end{table}

\begin{table}[t]
	\caption{Continued. }\vspace{3mm}
	\centering
	\begin{tabular}{lcccccccccc}
		\hline
		Hel.J.D. &\textit{V} & err&$U - B$&err&$B -V$&err&$V - R$&err&$R - I$&err \\ \hline \hline
		2452356.95 &0.700 &0.031 &2.480 &0.095 &1.938 &0.008 &1.640 &0.023 &1.449 &0.031 \\
		2452366.95 &0.723 &0.039 &2.721 &0.057 &1.918 &0.005 &1.687 &0.020 &1.428 &0.008 \\
		2452519.26 &0.331 &0.009 &2.362 &0.030 &1.923 &0.006 &1.523 &0.014 &1.317 &0.012 \\
		2452519.28 &0.342 &0.009 &2.328 &0.030 &1.949 &0.006 &1.527 &0.014 &1.313 &0.012 \\
		2452537.29 &0.320 &0.015 &2.325 &0.031 &1.942 &0.011 &1.503 &0.018 &1.329 &0.015 \\
		2452558.23 &0.334 &0.018 &2.370 &0.024 &1.967 &0.012 &1.502 &0.015 &1.313 &0.006 \\
		2452560.24 &0.374 &0.013 &2.324 &0.023 &1.970 &0.013 &1.521 &0.017 &1.316 &0.016 \\
\hline
	\end{tabular}\label{tabA2-02}
\end{table}

\begin{table}[t]
	\caption{Continued. }\vspace{3mm}
	\centering
	\begin{tabular}{lcccccccccc}
		\hline
		Hel.J.D. &\textit{V} & err&$U - B$&err&$B -V$&err&$V - R$&err&$R - I$&err \\ \hline \hline
		2453047.95 &0.470 &0.025 &2.709 &0.045 &2.018 &0.018 &1.575 &0.008 &1.307 &0.007 \\
		2453050.96 &0.470 &0.008 &2.501 &0.033 &1.987 &0.012 &1.542 &0.019 &1.381 &0.011 \\
		2453061.07 &0.488 &0.014 &2.524 &0.051 &2.004 &0.020 &1.588 &0.022 &1.408 &0.017 \\
		2453071.95 &0.524 &0.009 &2.580 &0.040 &1.961 &0.009 &1.599 &0.018 &1.370 &0.016 \\
		2453077.96 &0.539 &0.010 &2.417 &0.036 &1.976 &0.014 &1.560 &0.021 &1.324 &0.029 \\
		2453092.95 &0.550 &0.009 &2.554 &0.028 &1.975 &0.010 &1.613 &0.013 &1.414 &0.011 \\
		2453100.96 &0.548 &0.010 &2.631 &0.022 &2.008 &0.013 &1.617 &0.015 &1.388 &0.006 \\
\hline
	\end{tabular}\label{tabA2-03}
\end{table}

\begin{table}[t]
	\caption{Continued. }\vspace{3mm}
	\centering
	\begin{tabular}{lcccccccccc}
		\hline
		Hel.J.D. &\textit{V} & err&$U - B$&err&$B -V$&err&$V - R$&err&$R - I$&err \\ \hline \hline
		2453981.29 &0.894 &0.026 &2.241 &0.035 &1.919 &0.018 &1.674 &0.020 &1.563 &0.034 \\
		2454016.24 &0.925 &0.015 &2.327 &0.031 &1.931 &0.014 &1.702 &0.018 &1.604 &0.021 \\
		2454018.23 &0.993 &0.037 &2.528 &0.028 &1.995 &0.016 &1.724 &0.025 &1.638 &0.023 \\
		2454023.21 &0.929 &0.020 &2.551 &0.025 &1.961 &0.011 &1.728 &0.015 &1.640 &0.015 \\
		2454028.20 &0.865 &0.010 &      &      &      &      &      &      &      &      \\
		2454047.16 &0.747 &0.008 &2.509 &0.051 &1.963 &0.023 &1.685 &0.022 &1.454 &0.021 \\
		2454047.20 &0.734 &0.008 &2.528 &0.051 &1.958 &0.023 &1.665 &0.022 &1.456 &0.021 \\
\hline
	\end{tabular}\label{tabA2-04}
\end{table}

\begin{table}[t]
	\caption{Continued. }\vspace{3mm}
	\centering
	\begin{tabular}{lcccccccccc}
		\hline
		Hel.J.D. &\textit{V} & err&$U - B$&err&$B -V$&err&$V - R$&err&$R - I$&err \\ \hline \hline
		2454398.20 &0.967 &0.042 &2.469 &0.028 &1.938 &0.011 &1.713 &0.022 &1.580 &0.013 \\
		2454398.23 &0.934 &0.042 &2.416 &0.028 &1.920 &0.011 &1.709 &0.022 &1.597 &0.013 \\
		2454402.19 &0.979 &0.012 &2.196 &0.028 &1.939 &0.014 &1.687 &0.016 &1.650 &0.013 \\
		2454402.23 &0.984 &0.012 &2.184 &0.028 &1.936 &0.014 &1.701 &0.016 &1.632 &0.013 \\
		2454408.28 &0.934 &0.020 &2.349 &0.012 &1.925 &0.016 &1.654 &0.019 &1.580 &0.004 \\
		2454417.15 &1.016 &0.008 &2.270 &0.011 &1.932 &0.007 &1.727 &0.015 &1.603 &0.018 \\
		2454420.18 &0.990 &0.017 &2.324 &0.011 &1.926 &0.015 &1.733 &0.015 &1.599 &0.016 \\
\hline
	\end{tabular}\label{tabA2-05}
\end{table}

\begin{table}[t]
	\caption{Continued. }\vspace{3mm}
	\centering
	\begin{tabular}{lcccccccccc}
		\hline
		Hel.J.D. &\textit{V} & err&$U - B$&err&$B -V$&err&$V - R$&err&$R - I$&err \\ \hline \hline
		2454719.29 &0.701 &0.020 &2.191 &0.061 &1.941 &0.007 &1.635 &0.012 &1.484 &0.011 \\
		2454732.29 &0.720 &0.016 &2.111 &0.008 &1.933 &0.012 &1.583 &0.015 &1.488 &0.003 \\
		2454741.25 &0.796 &0.013 &2.286 &0.010 &1.963 &0.008 &1.660 &0.020 &1.519 &0.007 \\
		2454749.25 &0.870 &0.010 &2.223 &0.008 &1.959 &0.007 &1.637 &0.009 &1.526 &0.013 \\
		2454751.23 &0.807 &0.027 &2.305 &0.015 &1.923 &0.010 &1.628 &0.015 &1.503 &0.012 \\
		2454756.24 &0.794 &0.012 &2.369 &0.011 &1.948 &0.011 &1.668 &0.014 &1.478 &0.006 \\
		2454784.14 &0.949 &0.012 &2.246 &0.010 &1.922 &0.009 &1.685 &0.011 &1.579 &0.014 \\
\hline
	\end{tabular}\label{tabA2-06}
\end{table}

\begin{table}[t]
	\caption{Continued. }\vspace{3mm}
	\centering
	\begin{tabular}{lcccccccccc}
		\hline
		Hel.J.D. &\textit{V} & err&$U - B$&err&$B -V$&err&$V - R$&err&$R - I$&err \\ \hline \hline
		2455150.18 &0.361 &0.022 &2.239 &0.021 &1.919 &0.012 &1.537 &0.014 &1.340 &0.023 \\
		2455150.28 &0.383 &0.022 &2.229 &0.021 &1.939 &0.012 &1.514 &0.014 &1.322 &0.023 \\
		2455154.07 &0.328 &0.007 &      &      &      &      &      &      &      &      \\
		2455155.12 &0.366 &0.015 &2.349 &0.014 &1.941 &0.009 &1.517 &0.014 &1.339 &0.012 \\
		2455167.16 &0.392 &0.012 &2.339 &0.019 &1.962 &0.006 &1.536 &0.008 &1.294 &0.011 \\
		2455172.07 &0.355 &0.013 &2.385 &0.019 &1.974 &0.012 &1.512 &0.012 &1.325 &0.010 \\
		2455181.05 &0.327 &0.019 &      &      &      &      &1.508 &0.017 &1.289 &0.011 \\
\hline
	\end{tabular}\label{tabA2-07}
\end{table}

\begin{table}[t]
	\caption{Continued. }\vspace{3mm}
	\centering
	\begin{tabular}{lcccccccccc}
		\hline
		Hel.J.D. &\textit{V} & err&$U - B$&err&$B -V$&err&$V - R$&err&$R - I$&err \\ \hline \hline
		2455558.03 &0.365 &0.017 &2.384 &0.027 &1.950 &0.010 &1.528 &0.015 &1.328 &0.016 \\
		2455563.98 &0.414 &0.013 &2.352 &0.017 &1.927 &0.010 &1.510 &0.017 &1.380 &0.010 \\
		2455565.98 &0.421 &0.017 &      &      &      &      &1.554 &0.008 &      &      \\
		2455570.01 &0.415 &0.014 &2.366 &0.018 &1.943 &0.007 &1.551 &0.016 &1.351 &0.014 \\
		2455578.99 &0.435 &0.005 &2.405 &0.016 &1.942 &0.007 &1.518 &0.017 &1.344 &0.012 \\
		2455584.06 &0.458 &0.018 &2.343 &0.015 &1.953 &0.009 &1.566 &0.012 &1.387 &0.015 \\
		2455589.10 &0.490 &0.022 &2.260 &0.018 &1.952 &0.005 &1.559 &0.010 &1.321 &0.010 \\
\hline
	\end{tabular}\label{tabA2-08}
\end{table}

\begin{table}[t]
	\caption{Continued. }\vspace{3mm}
	\centering
	\begin{tabular}{lcccccccccc}
		\hline
		Hel.J.D. &\textit{V} & err&$U - B$&err&$B -V$&err&$V - R$&err&$R - I$&err \\ \hline \hline
		2456011.94 &0.505 &0.013 &      &      &      &      &1.564 &0.015 &      &      \\
		2456018.93 &0.663 &0.026 &2.418 &0.016 &1.988 &0.011 &1.593 &0.013 &1.406 &0.012 \\
		2456026.93 &0.634 &0.019 &      &      &      &      &1.627 &0.016 &      &      \\
		2456159.30 &0.460 &0.007 &      &      &      &      &1.552 &0.014 &      &      \\
		2456165.30 &0.459 &0.010 &2.192 &0.017 &2.020 &0.020 &1.524 &0.019 &1.335 &0.014 \\
		2456181.29 &0.451 &0.014 &2.222 &0.033 &1.976 &0.005 &1.547 &0.013 &1.369 &0.017 \\
		2456194.30 &0.516 &0.006 &      &      &      &      &      &      &      &      \\
\hline
	\end{tabular}\label{tabA2-09}
\end{table}

\begin{table}[t]
	\caption{Continued. }\vspace{3mm}
	\centering
	\begin{tabular}{lcccccccccc}
		\hline
		Hel.J.D. &\textit{V} & err&$U - B$&err&$B -V$&err&$V - R$&err&$R - I$&err \\ \hline \hline
		2456626.08 &0.544 &0.009 &      &      &1.978 &0.010 &1.551 &0.022 &1.408 &0.011 \\
		2456638.05 &0.548 &0.012 &      &      &      &      &1.586 &0.018 &      &      \\
		2456649.05 &0.556 &0.009 &2.272 &0.022 &1.951 &0.008 &1.567 &0.012 &1.410 &0.008 \\
		2456658.04 &0.534 &0.009 &      &      &      &      &1.563 &0.017 &      &      \\
		2456670.02 &0.501 &0.015 &      &      &      &      &1.583 &0.013 &      &      \\
		2456675.06 &0.477 &0.011 &2.235 &0.012 &1.968 &0.008 &1.551 &0.016 &1.371 &0.007 \\
		2456680.95 &0.470 &0.011 &      &      &      &      &1.562 &0.011 &      &      \\
\hline
	\end{tabular}\label{tabA2-10}
\end{table}

\begin{table}[t]
	\caption{Continued. }\vspace{3mm}
	\centering
	\begin{tabular}{lcccccccccc}
		\hline
		Hel.J.D. &\textit{V} & err&$U - B$&err&$B -V$&err&$V - R$&err&$R - I$&err \\ \hline \hline
		2457138.92 &0.350 &0.011 &      &      &      &      &      &      &      &      \\
		2457139.92 &0.377 &0.014 &      &      &      &      &      &      &      &      \\
		2457276.26 &0.659 &0.004 &      &      &      &      &      &      &      &      \\
		2457278.31 &0.665 &0.009 &      &      &      &      &      &      &      &      \\
		2457295.29 &0.703 &0.005 &      &      &      &      &      &      &      &      \\
		2457303.25 &0.712 &0.006 &2.100 &0.019 &1.922 &0.008 &1.633 &0.014 &1.466 &0.012 \\
		2457314.25 &0.724 &0.022 &      &      &1.961 &0.008 &1.680 &0.012 &1.501 &0.012 \\
\hline
	\end{tabular}\label{tabA2-11}
\end{table}

\begin{table}[t]
	\caption{Continued. }\vspace{3mm}
	\centering
	\begin{tabular}{lcccccccccc}
		\hline
		Hel.J.D. &\textit{V} & err&$U - B$&err&$B -V$&err&$V - R$&err&$R - I$&err \\ \hline \hline
		2457800.05 &0.377 &0.009 &      &      &      &      &      &      &      &      \\
		2457801.97 &0.389 &0.013 &      &      &      &      &1.511 &0.019 &      &      \\
		2457812.95 &0.285 &0.023 &      &      &      &      &1.486 &0.012 &      &      \\
		2457815.91 &0.336 &0.009 &      &      &      &      &1.526 &0.018 &      &      \\
		2457821.97 &0.326 &0.015 &2.364 &0.041 &2.000 &0.008 &1.498 &0.022 &1.279 &0.012 \\
		2457856.95 &0.292 &0.011 &      &      &      &      &      &      &      &      \\
		2457873.92 &0.257 &0.030 &      &      &      &      &      &      &      &      \\
\hline
	\end{tabular}\label{tabA2-12}
\end{table}

\begin{table}[t]
	\caption{Continued. }\vspace{3mm}
	\centering
	\begin{tabular}{lcccccccccc}
		\hline
		Hel.J.D. &\textit{V} & err&$U - B$&err&$B -V$&err&$V - R$&err&$R - I$&err \\ \hline \hline
		2458741.28 &0.625 &0.012 &      &      &1.986 &0.013 &1.644 &0.011 &1.423 &0.012 \\
		2458761.24 &0.620 &0.007 &1.963 &0.025 &1.899 &0.015 &1.617 &0.012 &1.457 &0.012 \\
		2458769.28 &0.640 &0.009 &      &      &      &      &1.618 &0.013 &      &      \\
		2458788.16 &0.766 &0.012 &      &      &      &      &1.657 &0.018 &      &      \\
		2458794.16 &0.803 &0.012 &1.916 &0.011 &1.944 &0.019 &1.622 &0.013 &1.577 &0.009 \\
		2458803.14 &0.860 &0.016 &2.041 &0.023 &1.880 &0.012 &1.714 &0.023 &1.576 &0.014 \\
		2458826.04 &1.150 &0.017 &      &      &1.881 &0.008 &1.746 &0.013 &      &      \\
\hline
	\end{tabular}\label{tabA2-13}
\end{table}

\begin{table}[t]
	\caption{Continued. }\vspace{3mm}
	\centering
	\begin{tabular}{lcccccccccc}
		\hline
		Hel.J.D. &\textit{V} & err&$U - B$&err&$B -V$&err&$V - R$&err&$R - I$&err \\ \hline \hline
		2459190.10 &0.755 &0.016 &2.110 &0.012 &1.967 &0.019 &1.648 &0.013 &1.500 &0.009 \\
		2459199.07 &0.747 &0.007 &2.294 &0.028 &1.992 &0.006 &1.626 &0.014 &1.520 &0.010 \\
		2459217.08 &0.639 &0.006 &2.245 &0.021 &1.998 &0.011 &1.610 &0.013 &1.461 &0.020 \\
		2459224.06 &0.632 &0.007 &      &      &      &      &1.637 &0.013 &      &      \\
		2459232.99 &0.573 &0.020 &2.223 &0.006 &1.987 &0.010 &1.582 &0.012 &1.432 &0.015 \\
		2459244.01 &0.584 &0.008 &2.277 &0.013 &2.002 &0.011 &1.598 &0.017 &1.418 &0.012 \\
		2459249.99 &0.591 &0.008 &2.342 &0.025 &1.986 &0.010 &1.630 &0.019 &1.421 &0.012 \\
\hline
	\end{tabular}\label{tabA2-14}
\end{table}

\end{document}